\documentclass[aps,reprint,superscriptaddress,amsmath,amssymb,prb,floatfix,showpacs]{revtex4-1}
\usepackage[utf8]{inputenc}
\usepackage{graphicx}

\usepackage{color}
\usepackage[usenames,dvipsnames]{xcolor}
\usepackage[colorlinks,bookmarks=false,citecolor=blue,linkcolor=black,urlcolor=blue]{hyperref}

\begin{document}
\title{
Order by disorder in a four flavor Mott-insulator on the fcc lattice
}

\author{P\'eter Sinkovicz}
 \affiliation{Institute for Solid State Physics and Optics, Wigner Research Centre for Physics, Hungarian Academy of Sciences, H-1525 Budapest, P.O.B. 49, Hungary}
\author{Gergely Szirmai}
 \affiliation{Institute for Solid State Physics and Optics, Wigner Research Centre for Physics, Hungarian Academy of Sciences, H-1525 Budapest, P.O.B. 49, Hungary}
\author{Karlo Penc}
 \affiliation{Institute for Solid State Physics and Optics, Wigner Research Centre for Physics, Hungarian Academy of Sciences, H-1525 Budapest, P.O.B. 49, Hungary}
\affiliation{MTA-BME Lend\"ulet Magneto-optical Spectroscopy Research Group, 1111 Budapest, Hungary}
\date{\today}

\begin{abstract}
The classical ground states of the SU(4) Heisenberg model on the face centered cubic lattice constitute a highly degenerate manifold.  
We explicitly construct all the classical ground states of the model. 
 To describe quantum fluctuations above these classical states, we apply linear flavor-wave theory. 
At zero temperature, the bosonic flavor waves select the simplest of these SU(4) symmetry breaking states, the four-sublattice ordered state defined by the cubic unit cell of the fcc lattice.
Due to geometrical constraints, flavor waves interact along specific planes only, thus rendering the system effectively two dimensional and forbidding ordering at finite temperatures. 
We argue that longer range interactions generated by quantum fluctuations can shift the transition to finite temperatures.
\end{abstract}

%
%
%
%
%

\maketitle

\section{Introduction}

The low energy physics of the Mott-insulating state in transition-metal oxides with orbital degeneracy is generally described by a Kugel-Khomskii model,\cite{KugelKhomskii_1982,brink2011} an extension of the Heisenberg model to include the coupling between the spin and orbital degrees of freedom between the ions. For electrons in the $e_g$ orbitals, the local Hilbert space is four dimensional, as next to the twofold spin degeneracy a twofold orbital degeneracy appears. The highest allowed symmetry in the Kugel-Khomskii model is SU(4),\cite{li1998su,yamashita1998,Kugel_PRB_2015} where all the spins and orbitals are treated equally, leading to the SU(4) Heisenberg model, with the Hamiltonian
\begin{equation}
\mathcal{H} = J \sum_{\left< \mathbf{r},\mathbf{r}' \right>} \mathcal{P}_{\mathbf{r},\mathbf{r}' } \, ,
\label{eq:Heisenber_proj}
\end{equation}
where $\mathcal{P}_{\mathbf{r},\mathbf{r}'}$ is the permutation operator exchanging the SU(4) flavors between the neighboring sites $\mathbf{r}$ and $\mathbf{r}'$. More generally, if the local Hilbert space is $N$-dimensional, the Hamiltonian~(\ref{eq:Heisenber_proj}) is SU($N$) symmetric, which, as a special case, includes the standard spin-1/2 SU(2) Heisenberg model. The exchange interaction in some $f$-electron materials, like the CeB$_6$,\cite{shiina1997} is also very close to the  SU(4) symmetry. 

A further -- and quite promising -- realization of SU($N$) symmetric Mott insulators are ultracold atoms on optical lattices\cite{lewenstein2012ultracold}, where an elaborate control over almost every single parameter of the model system is routinely utilized. The atoms are trapped optically; both the potential height and the lattice periodicity can be adjusted by tuning the amplitude, phase and wavelength of the lasers. Even the geometry of the lattice can be changed in situ.\cite{tarruell2012creating} Further advantage of such systems is that interaction between the atoms can be controlled in a wide range through the access of various scattering resonances.\cite{inouye1998observation,fedichev1996influence,petrov2001interatomic} In first experiments the Mott insulator was realized with a dilute gas sample of alkaline atoms loaded to an optical lattice.\cite{greiner2002quantum,jordens2008mott,greif2014short} Later, by trapping higher spin alkalies \cite{krauser2012coherent} and by cooling alkaline-earth-metal atoms to quantum degeneracy \cite{taie2010realization,desalvo2010degenerate} has opened the way to Mott insulators with higher spin atoms.\cite{taie2012su,scazza2014observation} As a result, one can hope that many antiferromagnets, either encountered in real materials, or proposed by theorists for academic interest can now be realized.\cite{gorshkov2010two,szirmai2011exotic,szirmai2011gauge,*sinkovicz2013spin,pinheiro2013xyz,cai2013pomeranchuk,cai2013quantum,cazalilla2014ultracold}

A large variety of ground states has been put forward for the SU(4) Heisenberg models. A quantum liquid with algebraically decaying  spin-spin correlations is formed in the one dimensional chain\cite{yamashita1998,Frischmuth1999} (which is Bethe ansatz solvable\cite{Sutherland1975}),  and there is strong evidence for a similar critical state in the hexagonal lattice.\cite{Corboz_honeycomb_2012} 
In ladder,\cite{van_den_bossche2001} checkerboard lattice,\cite{corboz2012} and honeycomb lattice with first and second neighbor exchange\cite{Lajko2013} the spins form SU(4) singlets or plaquettes, resulting in a translational symmetry-breaking states.
In the square lattice, the proposals range from plaquettes,\cite{Bossche2000,szirmai2011exotic} via liquid,\cite{fa2009} to a dimerized state.\cite{corboz2011} 
The SU(4) Heisenberg model in three-dimensional lattices is largely unexplored, with the exception of the cubic lattice, where a resonating plaquette state is suggested in Refs.~[\onlinecite{Pankov2007},\onlinecite{Xu2008}].  

In this paper, we study the ground state of the SU(4) symmetric Heisenberg antiferromagnet on a face centered cubic (fcc) lattice. Assuming an ordered, SU(4) symmetry breaking ground state, we first study the classical limit of the problem. We shall see, that the classical ground state is macroscopically degenerate and that the inclusion of quantum fluctuations in linear flavor-wave theory\cite{joshi1999elementary,zapf2014bose} reduces the degeneracy to the order of unity. The flavor wave excitations will have flat modes along one of the directions. In particular, the Goldstone mode \cite{goldstone1962broken} will have zero energy all along this line, which is the consequence of the classical degeneracy. According to the Mermin-Wagner-Hohenberg theorem, such a reduction of the effective dimension of the model, from 3 to 2, destroys long range order at finite temperatures.\cite{mermin1966absence,hohenberg1967existence} We show that the inclusion of the ubiquitous, but mostly negligible, next-nearest-neighbor interactions (originating either from exchange terms or generated by quantum fluctuations) couple the otherwise independent directions and stabilize a N\'eel ordered phase at finite temperature. 

The rest of the paper is organized as follows. In Sec.~\ref{sec:model}  we introduce the model first in terms of the SU(4) spin operators and later in terms of bosonic fields by applying the Schwinger representation. In Sec.~\ref{sec:Classical} the classical ordering, and in particular, its degeneracy is discussed in terms of the Schwinger boson mean fields. Quantum fluctuations are treated using flavor waves in Sec.~\ref{sec:quantum}, where we also show the selection of the ground state.
In Sec.~\ref{sec:NNN} we introduce an effective second neighbor interaction to describe the quantum effects at classical level. In Sec.~\ref{sec:correction} we explore the reduction of the ordered moments due to the quantum fluctuations at zero and finite temperatures. 
We summarize our results in Sec.~\ref{sec:sum}.

\section{The SU(4) Heisenberg model and its bosonic representation}
\label{sec:model}

Here we consider the SU(4) symmetric antiferromagnetic ($J>0$) Heisenberg model on the fcc lattice defined by the Hamiltonian
\begin{equation}
\mathcal{H} = J \sum_{\left< \mathbf{r},\mathbf{r}' \right>} \sum_{\alpha, \beta} S^\alpha_\beta(\mathbf{r}) S_\alpha^\beta(\mathbf{r}') \, .
\label{eq:Heisenberg}
\end{equation}
The first summation is over the nearest neighbors,  $S_\alpha^\beta(\mathbf{r})$ are the generators of the SU(4) Lie algebra at site $\mathbf{r}$, which follow the commutation relation~\cite{li1998su}
\begin{equation}
\left[ S^\alpha_\beta(\mathbf{r}), S^\mu_\nu(\mathbf{r}) \right] = \delta_{\alpha,\nu}S^\mu_\beta(\mathbf{r}) - \delta_{\beta, \mu}S^\alpha_\nu(\mathbf{r}) \,  ,
\label{eq:Lie_algebra}
\end{equation}
where $\alpha, \beta, \mu, \nu \in \{ A,B,C,D \}$. Generators on different sites commute. These 16 generators are the spin operators, which can be represented as $d_{\mathrm{IR}}\times d_{\mathrm{IR}}$ matrices, where $d_{\mathrm{IR}}$ is the dimension of the irreducible representation of the SU(4) algebra which spans the local Hilbert space. More precisely, since the trace $\sum_{\alpha} S^\alpha_\alpha(\mathbf{r})$ is proportional the identity operator, the number of nontrivial generators is only 15.

When the Heisenberg Hamiltonian \eqref{eq:Heisenberg} is realized as the low energy effective Hamiltonian of a Mott insulator with a single particle per site,\cite{auerbach1994interacting} the local Hilbert space is spanned by the states $\{A,B,C,D\}$, so that $S^\beta_\alpha|\alpha\rangle = |\beta\rangle$ and zero otherwise. The Hamiltonian (\ref{eq:Heisenberg}) then simplifies to Eq.~(\ref{eq:Heisenber_proj}), furthermore  
\begin{equation}
 \sum_{\alpha} S^\alpha_\alpha(\mathbf{r}) = \mathbb{I}_{4\times4} \;,
  \label{eq:saa}
\end{equation}
$\mathbb{I}_{4\times4}$ being the 4-dimensional unit matrix.
 The four states, we shall also refer to them as flavors or colors, are the four internal states of the particles and form the basis set of the 4-dimensional fundamental representation. These 4 states are the SU(4) analogs of the $\{\uparrow,\downarrow\}$ basis states of spin--1/2 representation, which is the fundamental representation of the SU(2) algebra.

The SU(4) Lie algebra, Eq.~(\ref{eq:Lie_algebra}), can be satisfied with the spin-operators of the following form \cite{auerbach1994interacting,arovas1988functional,zapf2014bose}
\begin{equation}
\label{eq:S_and_b}
S^\alpha_\beta(\mathbf{r}) = b_{\mathbf{r},\alpha}^\dagger b^{\phantom{\dagger}}_{\mathbf{r},\beta} \, ,
\end{equation}
where $b_{\mathbf{r},\alpha}^\dagger$ ($b^{\phantom{\dagger}}_{\mathbf{r},\beta}$) create (annihilate) a 
Schwinger boson with color $\alpha$ ($\beta$) on site $\mathbf{r}$. In this case we describe a Mott insulator with an integer number $M$ of bosonic particles on a site, so that 
\begin{equation}
\sum_\alpha b_{\mathbf{r},\alpha}^\dagger b^{\phantom{\dagger}}_{\mathbf{r},\alpha} = M  \, ,
\label{eq:constrait}
\end{equation}
which is equivalent to [c.f. Eq.~(\ref{eq:saa})] 
\begin{equation}
\sum_{\alpha} S^\alpha_\alpha(\mathbf{r}) = M  \mathbb{I}_{d_{\mathrm{IR}}\times d_{\mathrm{IR}}}\;.
\end{equation} 
On each site the 
\begin{equation}
d_{\mathrm{IR}}=\binom{M+3}{3} 
\end{equation}
dimensional fully symmetrical irreducible representation is realized, pictured by a Young tableau with a single row having $M$ boxes.\cite{arovas1988functional}
For a single boson per site, when $M$=1, we recover the $d_{\mathrm{IR}}=4$ fundamental representation (its Young tableau is a single box).
In the Schwinger boson representation the exchange interaction on the bonds takes the following form:
\begin{equation}
\sum_{\alpha, \beta} S^\alpha_\beta(\mathbf{r}) S_\alpha^\beta(\mathbf{r}') = \sum_{\alpha, \beta} b_{\mathbf{r},\alpha}^\dagger b_{\mathbf{r}',\beta}^\dagger b^{\phantom{\dagger}}_{\mathbf{r},\beta} b^{\phantom{\dagger}}_{\mathbf{r}',\alpha} \, .
\label{eq:permutation}
\end{equation}

In the equations above, $M$ is a parameter, allowing for a controlled treatment of fluctuations above the classical state, which is exact for $M\to\infty$. The role of fluctuations will be analyzed in flavor-wave theory as an expansion in $1/M$, just like the spin-wave theory is a $1/S$ expansion around the classical, $S\to \infty$ limit. We will keep the parameter $M$ explicit in the following to keep track of the order of the expansion. Our goal is to describe an ordered Mott insulator with a single particle (boson or fermion) per site, which corresponds to $M=1$.

\section{The classical ground states}
\label{sec:Classical}

\subsection{Classical states}

When thermal and quantum fluctuations are negligible, i.e. in the classical limit, boson operators are characterized by their mean values
\begin{subequations}
\label{eq:bean_values}
\begin{eqnarray}
\langle b^{\phantom{\dagger}}_{\mathbf{r},\alpha} \rangle = \sqrt{M} \xi^{\phantom{*}}_{\mathbf{r},\alpha} \, , \\
\langle b^{\dagger}_{\mathbf{r},\alpha} \rangle  = \sqrt{M} \xi^{*}_{\mathbf{r},\alpha} \, ,
\end{eqnarray}
\end{subequations}
where the classical field $\bm{\xi}_{\mathbf{r}}$ is a complex 4-dimensional unit vector at site $\mathbf{r}$ with components $\xi_{\mathbf{r},\alpha}$. The classical value of the spin operators \eqref{eq:S_and_b} are therefore represented by $S^\alpha_\beta(\mathbf{r})=\xi_{\mathbf{r},\alpha}^{*}\xi^{}_{\mathbf{r},\beta}$. In this limit they are no longer operators, instead they are $4\times4$ Hermitian matrices, furthermore they are rank-1 projectors. In comparison to the SU(2) case, there, a classical state is described, on similar grounds, by two-component bosons, and therefore, the classical field is a two component, complex, unit vector. The spin matrix at each site is a $2\times2$ Hermitian matrix, which is a rank-1 projector too. Such a projector is described by 2 real numbers, what we can identify as a unit vector in the 3D world, pointing to the direction of local magnetisation. In the SU(4) case the spin operators are characterised by 6 independent real numbers.

 \begin{figure}[tb!]
\includegraphics[width=0.7 \columnwidth]{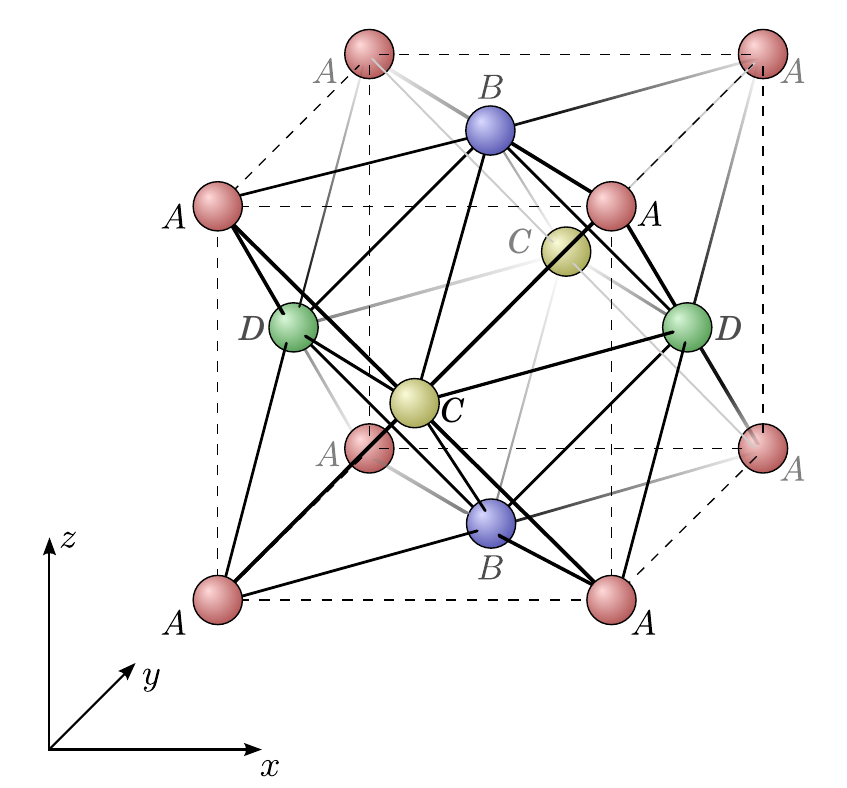}
\caption{(color online) The cubic unit cell of the four-sublattice ordered state. The coloring of the sites on the figure corresponds to the $\vartheta_z = 0$ case in Eq.~(\ref{eq:vartheta_Z}).
 \label{fig:trivial}
 }
\end{figure}

Starting from Eq.~\eqref{eq:Heisenberg}, the classical energy is obtained by replacing the bosonic operators in Eq.~(\ref{eq:permutation}) with their mean values given by Eqs.~(\ref{eq:bean_values}), 
\begin{equation}
E^{(0)} = J M^2 \sum_{\left< \mathbf{r},\mathbf{r}' \right>} \left| \sum_{\alpha} \xi_{\mathbf{r},\alpha}^* \xi^{\phantom{*}}_{\mathbf{r}',\alpha} \right|^2.
\label{eq:H_exp}
\end{equation}
The classical ground state configuration is determined by minimizing $E^{(0)}$ with respect to the classical fields. Since it is a sum of squares, the energy of the classical ground state is bounded from below by zero. Zero energy is realized for configurations with orthogonal $\bm{\xi}$ vectors on neighboring sites.  The simplest classical ground state can be found naturally in the cubic unit cell of the fcc lattice: we can choose four orthogonal colors on the four sublattices, as shown in Fig.~\ref{fig:trivial}. However, this is not the only choice.
In the following, we are going to show that the classical ground state for the fcc lattice is highly degenerate, and the problem is in fact underconstrained.

\subsection{The degeneracy of the classical ground states}
The most general classical configurations consist of two sublattice-ordered $\{100\}$ layers as follows:  let us consider a single plane of the lattice, for example the $z=0$ $(001)$ plane (the normal vector of this plane points to the $z$ direction, e.g. the lowest horizontal plane in Fig.~\ref{fig:trivial}). The sites in this plane form a bipartite square lattice and can be colored with two colors, i.e. we can choose two orthogonal $\bm{\xi}$ vectors as $(\cos\vartheta_0,\sin\vartheta_0,0,0)$ and $(-\sin\vartheta_0,\cos\vartheta_0,0,0)$.
On the following $z=1/2$ plane (we set the lengths of the sides of the cubic unit cell to 1), which is also a square lattice, we can select the two $\bm{\xi}$ vectors as $(0,0,\cos\vartheta_{1/2},\sin\vartheta_{1/2})$ and $(0,0,-\sin\vartheta_{1/2},\cos\vartheta_{1/2})$ on the two sublattices, respectively. On the $z=1$ layer the allowed $\bm{\xi}$ vectors are $(\cos\vartheta_1,\sin\vartheta_1,0,0)$ and $(-\sin\vartheta_1,\cos\vartheta_1,0,0)$ -- this is the same as on the plane $z=0$, except for the choice of $\vartheta$. The rule is actually simple: by defining  $\bm{\xi}_{A}=(1,0,0,0)$, $\bm{\xi}_{B}=(0,1,0,0)$,  $\bm{\xi}_{C}=(0,0,1,0)$, and $\bm{\xi}_{D}=(0,0,0,1)$, we may choose two arbitrary, orthogonal linear combinations of $\bm{\xi}_{A}$ and $\bm{\xi}_{B}$ on layers with integer $z$, while on layers with half-integer $z$ we choose orthogonal combinations of $\bm{\xi}_{C}$ and $\bm{\xi}_{D}$. This state can be characterized by the direction of the plane -- in this case $(001)$, and by the ordered set of the 
\begin{equation}
\{\vartheta_z\} = (\dots,\vartheta_{-1},\vartheta_{-1/2},\vartheta_0,\vartheta_{1/2},\vartheta_1,\vartheta_{3/2},\ldots)
\label{eq:vartheta_Z}
\end{equation}
values. Clearly, we may have chosen a different plane -- the $(100)$, the $(010)$, and the $(001)$ are crystallographically equivalent. 
  The particular choice of the classical fields on the first two, $z=0$ and $z=1/2$, planes is without loss of generality, since there is always a suitable global SU(4) transformation which can rotate a state into the above form.
  Setting all the $\vartheta_z$ in Eq.~(\ref{eq:vartheta_Z}) to 0 corresponds to the four-sublattice-ordered state shown in Fig.~\ref{fig:trivial}: all the three $\{100\}$ planes  are two sublattice ordered.

\begin{figure}[tb!]
\includegraphics[width=1 \columnwidth]{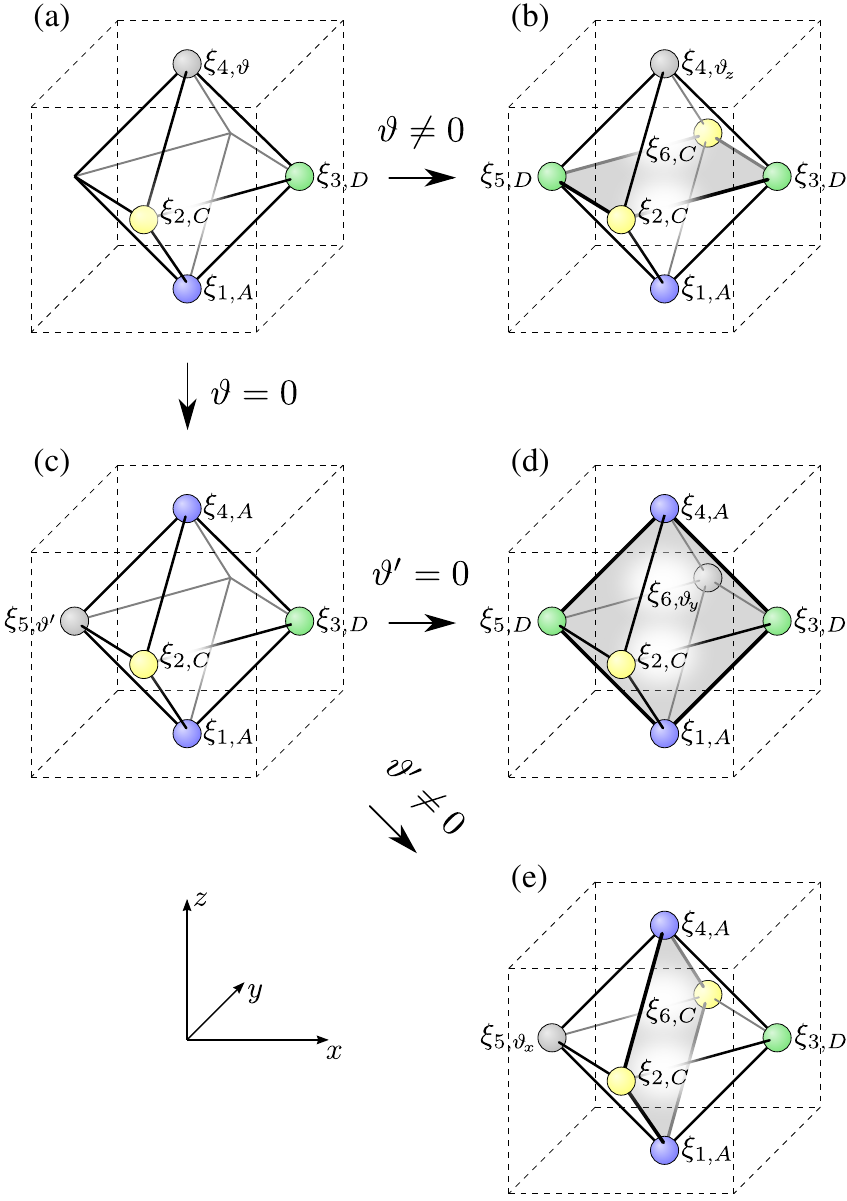}
\caption{(color online) A selection tree illustrating the degeneracy of the classical configurations on an octahedron. The figure shows how the classical fields on the sites of an octahedron formed by the face centers of the unit cell can be chosen one by one, in a way to minimize $E^{(0)}$ (see the text for details). The gray planes denote the planes colored by two flavors only.
\label{fig:classical_octahedron} }
\end{figure}

Now, let us show below that these are the only possible classical  ground state configurations. For that purpose, let us consider the octahedron shown in Fig.~\ref{fig:classical_octahedron}, constructed from the face centers of the cubic unit cell.
On sites 1, 2, and 3 (a triangle) the global SU(4) rotation allows us to select the classical fields to point to the directions $\bm{\xi}_A=(1,0,0,0)$, $\bm{\xi}_C=(0,0,1,0)$, and $\bm{\xi}_D=(0,0,0,1)$, respectively. On site 4, which is not connected to site 1, we can choose the classical field as a linear combination of $\bm{\xi}_A$ and $\bm{\xi}_B$, namely $\bm{\xi}_{4,\vartheta} = (\cos\vartheta,\sin\vartheta,0,0)$ [see Fig.~\ref{fig:classical_octahedron}(a)]. On one hand, we may select $\vartheta\neq 0$, then the classical fields on site 5 and 6 are fully determined, as shown in Fig.~\ref{fig:classical_octahedron}(b). On the other hand, if $\vartheta = 0$, then the classical field on site 5 can be chosen as $\bm{\xi}_{5,\vartheta'} = (0, \cos\vartheta', 0, \sin\vartheta')$, orthogonal to the vectors on the neighboring sites [see Fig.~\ref{fig:classical_octahedron}(c)]. 
Now, depending on whether we choose $\vartheta' = 0$ or not, we end up with the two possible configurations shown in Fig.~\ref{fig:classical_octahedron}(d) and Fig.~\ref{fig:classical_octahedron}(e). To summarize, starting from a given configuration on sites 1,2, and 3, we find three classes of solutions, where the $\bm{\xi}$ on one of the sites 4, 5, or 6 is not fully determined. Once the colors on the octahedron are decided, the  classical fields on the corners of the cube in Fig.~\ref{fig:classical_octahedron}(b) are determined too (the corner of the cube and the three nearest-neighbor face centers form a tetrahedron, and there is no coloring freedom for the complete graph of four sites). 
Similar considerations hold for Figs.~\ref{fig:classical_octahedron}(d) and (e). All possible outcomes (b), (d) and (e) have one thing in common: In each of them we can find four sites on a square which are coloured by two colours only, and which define a plane.
In Fig.~\ref{fig:classical_octahedron}(b), this plane is defined by the sites $2,3,6,5$, while on outcome (d), the plane is defined by sites $1,3,4,5$, and finally, on outcome (e), the plane is defined by sites $1,6,4,2$. These four sites form the basis of the two-sublattice ordered planes of the previous paragraph, when the fcc lattice is gradually built from the cubic cells each containing an octahedron and the corner points surrounding it.

\subsection{The helical state}
\label{subsec:classical_helical}
A particularly simple choice of the parameter set $\lbrace\vartheta_z\rbrace$ corresponds to the helical states,
defined by a $\vartheta_z$ varying linearly with the position of the plane along the $z$ axis,  $\vartheta_z = z\,\vartheta$ in Eq.~(\ref{eq:vartheta_Z}).
The classical field of the helical state can be expressed in the following recursive form:
\begin{equation}
\bm{\xi}_{\mathbf{r}+\bm{\delta}}= \mathbf{R}(-\vartheta/2)\, \bm{\xi}_{\mathbf{r}},
 \label{eq:xi_recursive}
\end{equation}
where  the matrix 
\begin{eqnarray}
\mathbf{R}(\vartheta) = 
	\left[
		\begin{array}{cccc}
			0 & 0 & \cos \vartheta & - \sin \vartheta \\
			0 & 0 & \sin \vartheta & \cos \vartheta \\
			\cos \vartheta & - \sin \vartheta & 0 & 0 \\
			\sin \vartheta & \cos \vartheta & 0 & 0\\
		\end{array}
	\right]
\label{eq:R}
\end{eqnarray}
transforms the fields from one layer to the following one translated by the vector $\bm{\delta} = \left(\frac{1}{2},0,\frac{1}{2} \right)$. Furthermore, we set the classical fields on a specific, reference plane, say the $z=0$ layer. On this layer we use a two sublattice ansatz, where the two sublattices $\Lambda_A$ and $\Lambda_B$ alternate in a checkerboard manner. In this plane, we choose $\bm{\xi}_\mathbf{r}=\bm{\xi}_A$, if $\mathbf{r}\in\Lambda_A$, and $\bm{\xi}_\mathbf{r}=\bm{\xi}_B$, if $\mathbf{r}\in\Lambda_B$. The classical fields on the other planes can be calculated from this reference plane with the help of the recursion \eqref{eq:xi_recursive}.

Now, it is convenient to extend the sublattices $\Lambda_A$ and $\Lambda_B$ to the whole fcc lattice: the sites that can be reached from $\Lambda_A$ in the $z=0$ plane by translations with $\bm{\delta}$ belong to sublattice $\Lambda_A$. Sublattice $\Lambda_B$ can be defined similarly. This way, $\Lambda_A$ consists of the lattice points $\mathbf{r}_i=(m_x,m_y,m_z)$ and $(m_x+1/2,m_y,m_z+1/2)$, while sites with coordinates $(m_x+1/2,m_y+1/2,m_z)$ and $(m_x,m_y+1/2,m_z+1/2)$ belong to $\Lambda_B$, where $m_x$, $m_y$, and $m_z$ are integers. 

Our simple recursion for the helical state, Eq. \eqref{eq:xi_recursive}, allows us to express the classical field explicitly at every site by raising the recursion matrix to the power corresponding to the distance of the site from the $z=0$ layer.  That is, 
\begin{align}
\label{eq:rotatingfield}
\bm{\xi}_{(x,y,n/2)} 
&=\mathbf{R}^n(-\vartheta/2)\cdot  \bm{\xi}_{(x-n/2,y-n/2,0)} \nonumber\\
& = \mathbf{R}^n(-\vartheta/2)\cdot  \bm{\tilde \xi}_{(x,y,n/2)},
\end{align}
where $\tilde{\bm{\xi}}_{\mathbf{r}}=\bm{\xi}_A$ for $\mathbf{r}\in\Lambda_A$ and $\tilde{\bm{\xi}}_{\mathbf{r}}=\bm{\xi}_B$ for the sites belonging to sublattice $\Lambda_B$.

\begin{figure*}[bt]
\includegraphics[width=1.9 \columnwidth]{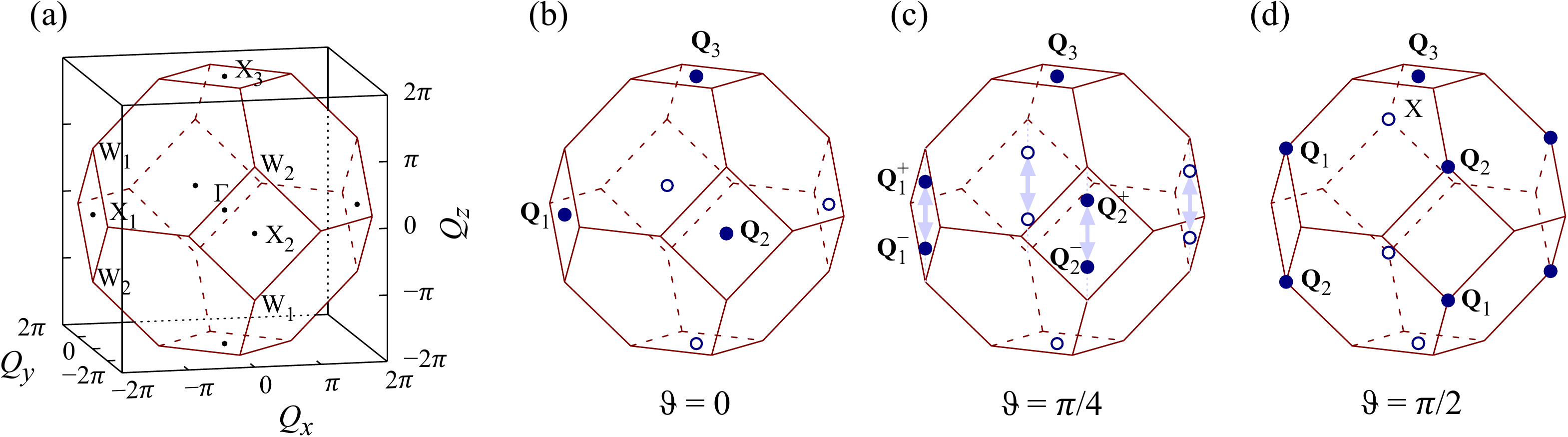}
\caption{(color online) Ordering vectors of the helical state in the Brillouin zone for three different values of $\vartheta$. (a) The Brillouin zone and high symmetry points of the fcc lattice. (b) The $\vartheta=0$ (the four-sublattice ordered) case: the ordering vectors $\mathbf{Q}_j$ ($j=1,2,3$) are located at the three distinct X high symmetry points. (c) The $\vartheta=\pi/4$ case: two of the ordering vectors $\mathbf{Q}_i$ split into $\mathbf{Q}_i^{+}$ and $\mathbf{Q}_i^{-}$ ($i=1,2$). These roots move towards the closest $W$ points as $\vartheta$ increases; (d) At $\vartheta=\pi/2$, the ordering wave vectors merge again at the $W_1$ and $W_2$ points, but now $\mathbf{Q}_{1} := \mathbf{Q}_1^+ = \mathbf{Q}_2^-$ and $\mathbf{Q}_2 := \mathbf{Q}_1^- = \mathbf{Q}_2^+$.
\label{fig:ordering_Q} }
\end{figure*}

In general, the structure is  incommensurate with the lattice spacing, except for special values of $\vartheta$. These include the $\vartheta = 0$ case, where $\mathbf{R}^2(\vartheta)$ is the identity matrix, and we recover the four--sublattice state shown in Fig.~\ref{fig:trivial}. The other special case is for $\vartheta=\pi/2$, with a periodicity $\Delta z=2$: two--sublattice ordered square lattice planes of pure $\bm{\xi}_A$ and $\bm{\xi}_B$ alternate with layers of pure $\bm{\xi}_C$ and $\bm{\xi}_D$ such that every fourth layer is identical. Two layers above $\bm{\xi}_A$ we find a $\bm{\xi}_B$; consequently, two layers above a $\bm{\xi}_C$ there is a $\bm{\xi}_D$ and vice versa. This state is discussed in more detail in Appendix~\ref{app:pi_2_case}.

\subsection{Classical spin-spin correlation function in the helical state}
\label{sec:correlation}

The order of the helical state can be characterized with the help of the so-called \emph{ordering wave vectors},\cite{auerbach1994interacting} which are the positions of the peaks of the static structure factor $S(\mathbf{q})$.\cite{auerbach1994interacting,altland2010condensed} The static structure factor is the Fourier transform of the equal time spin-spin correlation function between spins on sites $\mathbf{0}=(0,0,0)$ and $\mathbf{r}$,
\begin{equation}
S(\mathbf{r}) =  \sum_{\alpha,\beta} \left\langle
\left[ S_{\alpha}^\beta (\mathbf{0}) -\frac{M}{4}\delta_{\alpha,\beta}\right]
\left[ S_\beta^\alpha (\mathbf{r})  -\frac{M}{4}\delta_{\alpha,\beta}\right]
 \right\rangle ,
\end{equation}
where the trace of the SU(4) generators is subtracted. Using Eqs.~(\ref{eq:S_and_b}) and (\ref{eq:bean_values}), we take the expectation value in the classical ground state by completely neglecting fluctuations, arriving to 
\begin{equation}
S_{\text{cl}}(\mathbf{r}) 
 = M^2 \left( \Bigl| \sum_{\alpha} \xi_{\mathbf{0},\alpha}^* \xi^{\phantom{*}}_{\mathbf{r},\alpha} \Bigr|^2 - \frac{1}{4}\right)\;.
\label{eq:corr_cl}
\end{equation}
For the helical state, given by Eq.~(\ref{eq:rotatingfield}), it evaluates to
\begin{equation}
S_{\textrm{cl}}(\mathbf{r}) = 
\begin{cases} 
 (1 \!+\! 2 \cos 2 n_z \vartheta) \frac{M^2}{4}, 
 &\!\! \text{if } \mathbf{r}\! =\! (n_x,n_y,n_z); \\ 
(1 \!-\! 2 \cos 2 n_z \vartheta)\frac{M^2}{4},  
 &\!\! \text{if } \mathbf{r}\! =\! (n_x\!+\!\frac{1}{2},n_y\!+\!\frac{1}{2},n_z); \\
 -\frac{M^2}{4}, 
 &\!\! \text{if } \mathbf{r}\! =\! (n_x\!+\!\frac{1}{2},n_y,n_z\!+\!\frac{1}{2});\\
  -\frac{M^2}{4}, 
 &\!\! \text{if } \mathbf{r}\! =\! (n_x,n_y\!+\!\frac{1}{2},n_z\!+\!\frac{1}{2}). 
\end{cases} 
\label{eq:c_real_space}
\end{equation}
For the simple four-sublattice state, i.e. the  helical state with $\vartheta=0$, the correlation function simplifies to
\begin{equation}
S_{\textrm{cl}}^{(\vartheta = 0)}(\mathbf{r}) = 
\begin{cases} 
 \frac{3}{4}M^2\,, & \text{if } \mathbf{r} = (n_x,n_y,n_z)\,; \\ 
 -\frac{1}{4}M^2\,, & \text{otherwise}.
\end{cases} 
\end{equation}

The Fourier transform of the correlation function of the helical state, $S_{\text{cl}}(\mathbf{q}) = \sum_{\mathbf{r}} e^{-i \mathbf{r}\cdot \mathbf{q}} S_{\text{cl}}(\mathbf{r})$, consists of delta functions at the ordering wave vectors in the Brillouin zone [Fig.~\ref{fig:ordering_Q}(a)],
\begin{subequations}
\label{eqs:qpoints}
\begin{eqnarray}
\mathbf{Q}^{\pm}_1 &=& 2 \pi (1,0, \pm \vartheta / \pi) \, ,\\
\mathbf{Q}^{\pm}_2 &=& 2 \pi (0,1,\pm \vartheta / \pi ) \, ,\\
\mathbf{Q}_3 &=& 2 \pi (0,0,1) \, ,
\end{eqnarray}
\end{subequations}
and at the equivalent points in the bcc reciprocal lattice spanned by the reciprocal primitive vectors $2 \pi (1,1,1)$, $2 \pi (1,1,-1)$, and $2 \pi (1,-1,1)$. For the four sublattice case, the three $\mathbf{Q}$ vectors are located at the three distinct high symmetry points X. As $\vartheta$ becomes nonzero, both $\mathbf{Q}_1$ and $\mathbf{Q}_2$ split into two branches. These new points move towards the distinct W points, reaching them at $\vartheta=\pi/2$. The $W_1$ points (and similarly $W_2$)  are equivalent, differing only by a reciprocal primitive vector, therefore the two branches merge again at $\vartheta=\pi/2$, but now $\mathbf{Q}_1\equiv\mathbf{Q}_1^+=\mathbf{Q}_2^-$ ($W_1$ point), and $\mathbf{Q}_2\equiv\mathbf{Q}_2^+=\mathbf{Q}_1^-$ ($W_2$ point). See Fig.~\ref{fig:ordering_Q} for illustration.

The nonvanishing Fourier components of the correlation function are all equal for the $\vartheta=0$ and $\vartheta=\pi/2$ cases, with
\begin{equation}
S(\mathbf{Q}_1) =  S(\mathbf{Q}_2) = S(\mathbf{Q}_3) =  \frac{M}{4} 
\,,
\end{equation}
while for an arbitrary helical state
\begin{align}
S(\mathbf{Q}^{\pm}_1) =  S(\mathbf{Q}^\pm_2) = \frac{M}{8} \,,\quad
S(\mathbf{Q}_3) = \frac{M}{4} 
\,.
\end{align}
 These  can be identified as Bragg peaks in a corresponding scattering experiment.\cite{auerbach1994interacting,altland2010condensed}

\section{Quantum corrections from linear flavor-wave theory}
\label{sec:quantum}

We have seen in the previous section that the classical energy, Eq.~\eqref{eq:H_exp}, takes its minimum value on a highly degenerate set of configurations. In this section we show that the inclusion of quantum fluctuations lifts the degeneracy and selects a specific long-range-ordered state through the \emph{order-by-disorder} mechanism.\cite{shender1982,henley1989ordering}

For simplicity, we construct the linear flavor-wave theory over the helical state of Sec.~\ref{subsec:classical_helical}. That is, we look for quantum fluctuations over the classical configurations given by Eq. \eqref{eq:rotatingfield}. Our calculation is in the spirit of Ref.~[\onlinecite{toth2010}], where the degeneracy of the SU(3) Heisenberg model on the square lattice was treated. 

Using Eq.~\eqref{eq:R} as a canonical transformation, we introduce new boson operators as 
\begin{equation}
\label{eq:cantr}
\tilde{b}_{(x,y,n/2),\alpha}=\sum_\beta\left[\mathbf{R}^{n}(\vartheta/2)\right]_{\alpha\beta}b_{(x,y,n/2),\beta} \, .
\end{equation}
Since the mean value $\langle\tilde{b}_{\mathbf{r},\alpha}\rangle =\sqrt{M} \tilde\xi_{\mathbf{r},\alpha}$, the definition \eqref{eq:cantr} reduces to Eq.~\eqref{eq:rotatingfield} in the classical limit. The classical field $\tilde\xi_{\mathbf{r},\alpha}$ has a single nonzero component, which is equal to $\bm{\xi}_A$, $\bm{\xi}_B$ for $\mathbf{r}\in\Lambda_A,\Lambda_B$, respectively.
This canonical transformation allows us to replace  the transformed  Schwinger bosons $\tilde{\mathbf{b}}$ with Holstein-Primakoff bosons $\mathbf{c}$:\cite{joshi1999elementary}
\begin{subequations}
\label{eqs:HP}
\begin{equation}
\tilde{b}_{\mathbf{r},\alpha} = 
	\left\{
	\begin{array}{cc}
	c_{\mathbf{r},\alpha}, & \textrm{ if $\mathbf{r} \not\in  \Lambda_\alpha $},  \\
	&\\
	\sqrt{M-\mu_{\mathbf{r},\alpha}}, & \textrm{ if $\mathbf{r} \in  \Lambda_\alpha $},
	\end{array}
	\right.
\label{eq:cond_op}
\end{equation}
where
\begin{equation}
\label{eq:spinred}
\mu_{\mathbf{r},\alpha} = \sum_{\beta \neq \alpha} c^\dagger_{\mathbf{r},\beta} c_{\mathbf{r},\beta} \, .
\end{equation}
\end{subequations}
The Holstein-Primakoff bosons describe transverse fluctuations around the classical ordering. The amount of these fluctuation is measured by the occupation number $\mu_{\mathbf{r},\alpha}$. This quantity is also referred to as the spin reduction. 

\subsection{Calculation of the zero point energy}

We calculate the $\vartheta$ dependence of the quantum correction to the classical energy first by expanding Eq.~(\ref{eq:cond_op}) in powers of $1/M$, then through Eqs. \eqref {eq:cantr}  and \eqref{eq:permutation} we also expand the Hamiltonian \eqref{eq:Heisenberg}. The first term in the Hamiltonian is the classical energy $E^{(0)}$, proportional to $M^2$ (see Eq. \eqref{eq:H_exp}). It vanishes, because the classical ground state energy is zero. The second term $\propto M^{3/2}$ vanishes since the state is a local minimum of the classical energy. Therefore, the leading term is proportional to $M$, which is quadratic in the boson operators. The diagonalization of this quadratic term with a Bogoliubov transformation allows us to calculate the ground state energy resulting from fluctuations.

To progress further, we introduce a unit cell containing two sites, one from each sublattices $\Lambda_A$ and $\Lambda_B$. The position of the unit cell is given by 
$\mathbf{r} = (m_x,m_y,m_z)$ in the planes with integer $z$-coordinate and 
$\mathbf{r}=(m_x+\frac{1}{2},m_y,m_z+\frac{1}{2})$
 in the planes with half-integer $z$-coordinate. The unit cell contains one  site from sublattice $\Lambda_A$ (with coordinate $\mathbf{r}$), and another one from sublattice $\Lambda_B$ with coordinate $\mathbf{r}+\bm{\delta'}$, where $\bm{\delta'} = \left(\frac{1}{2},\frac{1}{2},0 \right)$. A relatively simple notation can be obtained by grouping the 6 Holstein-Primakoff bosons of the unit cell (3 for each site) into the formal vector
\begin{equation}
\mathbf{\Phi}_{\mathbf{r}} = (c_{\mathbf{r},B}, c_{\mathbf{r},C}, c_{\mathbf{r},D}, c_{\mathbf{r}+\bm{\delta'},A}, c_{\mathbf{r}+\bm{\delta'},C},  c_{\mathbf{r}+\bm{\delta'},D})^T,
\end{equation}
and its Fourier-transform
\begin{equation}
\mathbf{\Phi}_{q} = \sqrt{\frac{2}{N_s}} \sum_{\mathbf{r}} \mathbf{\Phi}_{\mathbf{r}}\,e^{- i \mathbf{q}\cdot\mathbf{r}} \, ,
\label{eq:fourier}
\end{equation}
where $N_s$ denotes the number of lattice sites.

In the semiclassical expansion we use the approximate form (up to the order of $M^{-3/2}$)
\begin{equation}
\sqrt{M-\mu_{\mathbf{r},\alpha}} \approx \sqrt{M} \left( 1-  \frac{\mu_{\mathbf{r},\alpha}}{2M} \right)\, ,
\label{eq:aprox}
\end{equation}
and we substitute it into the Hamiltonian (\ref{eq:Heisenber_proj}), which to leading order in $M$ gives
\begin{align}
\mathcal{H}^{(2)} = J M \sum_{\mathbf{q} \in BZ} \left[ 4 \, \mathbf{\Phi}_{\mathbf{q}}^{\dagger} \mathbf{\Phi}_{\mathbf{q}} + 
 2\left(  \mathbf{\Phi}_{\mathbf{q}}^T \mathbf{B}^*(\mathbf{q})  \mathbf{\Phi}_{-\mathbf{q}} + \text{h.c.} \right)  \right]  \, ,
 \label{eq:secondHAM}
\end{align}
where the matrix $\mathbf{B}(\mathbf{q})$ is
\begin{widetext}
\begin{eqnarray}
	 \mathbf{B}(\mathbf{q}) = \left[ 
		 \begin{array}{cccccc}
	 		- c_x c_y & 0 & 0 & 0 & 0 & 0 \\
	 		0 & c_y c_z c_\vartheta^2 & -i c_y s_z c_\vartheta s_\vartheta & 0 & -i c_x s_z c_\vartheta s_\vartheta & c_x c_z s_\vartheta^2 \\
	 		0 & i c_y s_z c_\vartheta s_\vartheta & - c_y c_z s_\vartheta^2 & 0 & - c_x c_z c_\vartheta^2 & -i c_x s_z c_\vartheta s_\vartheta \\
	 		0 & 0 & 0 & - c_x c_y & 0 & 0 \\
	 		0 & i c_x s_z c_\vartheta s_\vartheta & - c_x c_z c_\vartheta^2 & 0 & - c_y c_z s_\vartheta^2 & i c_y s_z c_\vartheta s_\vartheta \\
	 		0 & c_x c_z s_\vartheta^2 & i c_x s_z c_\vartheta s_\vartheta & 0 & -i c_y s_z c_\vartheta s_\vartheta & c_y c_z c_\vartheta^2 \\
	 	\end{array} 
	 \right] .
\end{eqnarray}
\end{widetext}
Above, we introduced the following shorthand notations: $c_\nu = \cos(q_\nu/2)$ and $s_\nu = \sin(q_\nu/2)$ for $\nu=x,y,z$, furthermore $c_\vartheta=\cos(\vartheta/2)$ and $s_\vartheta=\sin(\vartheta/2)$. 

The Hamiltonian (\ref{eq:secondHAM}) is quadratic in the boson operators, so it can be diagonalized by choosing a new set of operators with a Bogoliubov transformation
\begin{subequations}
\begin{eqnarray}
\tilde{\mathbf{\Phi}}_\mathbf{q} &=& \mathbf{U}(\mathbf{q})^\dagger \mathbf{\Phi}_\mathbf{q}  + \mathbf{V}^\dagger(\mathbf{q}) \mathbf{\Phi}^\dagger_{-\mathbf{q}} \, , \\
\tilde{\mathbf{\Phi}}_{-\mathbf{q}}^\dagger &=& \mathbf{U}^T(\mathbf{q}) \mathbf{\Phi}^\dagger_{-\mathbf{q}}  +  \mathbf{V}^T(\mathbf{q}) \mathbf{\Phi}_{\mathbf{q}} \, .
\end{eqnarray}
\end{subequations}
The transformation matrices are composed of the eigenvectors of the Bogoliubov matrix
\begin{equation}
\mathbf{M}(\mathbf{q}) =  4 J M \left[
 	\begin{array}{cc}
	\mathbb{I}_{6 \times 6} &  \mathbf{B}(\mathbf{q}) \\
	- \mathbf{B}^*(\mathbf{q}) &  - \mathbb{I}_{6 \times 6}
 	\end{array}
 	\right] \, ,
 	\label{eq:M_mtx}
\end{equation}
where $\mathbb{I}_{6\times6}$ is the 6-dimensional identity matrix. The  Bogoliubov matrix $\mathbf{M}$  defines the dynamics of the boson vector $(\mathbf{\Phi}^{\phantom{\dagger}}_\mathbf{q},\mathbf{\Phi}_{-\mathbf{q}}^\dagger)$,
\begin{equation}
\label{eq:eqmobogo}
i \hbar \,  \frac{\partial}{\partial t} \left( \begin{array}{cc} \mathbf{\Phi}_\mathbf{q}  \\ \mathbf{\Phi}_{-\mathbf{q}}^\dagger \end{array} \right) = \mathbf{M}(\mathbf{q}) \left( \begin{array}{cc} \mathbf{\Phi}_\mathbf{q}  \\ \mathbf{\Phi}_{-\mathbf{q}}^\dagger \end{array} \right) \, .
\end{equation}
The characteristic equation for the eigenvalues and transformation matrices $\mathbf{U}$ and $\mathbf{V}$ is given by
\begin{equation}
\label{eq:bogodiag}
\sum_k M_{j,k}(\mathbf{q}) T_{k,i}(\mathbf{q}) = \omega_i T_{j,i}(\mathbf{q})  \, ,
\end{equation}
where the transformation matrices are arranged in a block matrix form,
\begin{equation}
\mathbf{T}(\mathbf{q}) = \left( \begin{array}{rr} \mathbf{U}(\mathbf{q}) & - \mathbf{V}(\mathbf{q})  \\ - \mathbf{V}(\mathbf{q}) & \mathbf{U}(\mathbf{q}) \end{array} \right),
\end{equation}
and $i,j,k \in \{1,2,...,12 \}$. The eigenvalues are real and come in $\pm$ pairs. We order these eigenvalues, so that $\omega_1>\omega_2>\cdots > \omega_6 >0$ and $\omega_i=-\omega_{6+i}$. The eigenvectors are normalized with the requirement that the new operators $\tilde{\mathbf{\Phi}}$ fulfill boson commutation relations: \begin{subequations}
\begin{eqnarray}
\mathbf{U}(\mathbf{q}) \mathbf{U}^\dagger(\mathbf{q}) - \mathbf{V}(\mathbf{q}) \mathbf{V}^\dagger(\mathbf{q}) &=& \mathbb{I}_{6\times 6} \, , \\
\mathbf{V}^\dagger(\mathbf{q}) \mathbf{U}^*(\mathbf{q}) - \mathbf{U}^\dagger(\mathbf{q}) \mathbf{V}^*(\mathbf{q}) &=& \mathbb{O}_{6 \times 6} \, ,
\end{eqnarray}
\end{subequations}
where $\mathbb{O}_{6 \times 6}$ is the 6-dimensional zero matrix.

\begin{figure}[tb!]
\includegraphics[width=0.9 \columnwidth]{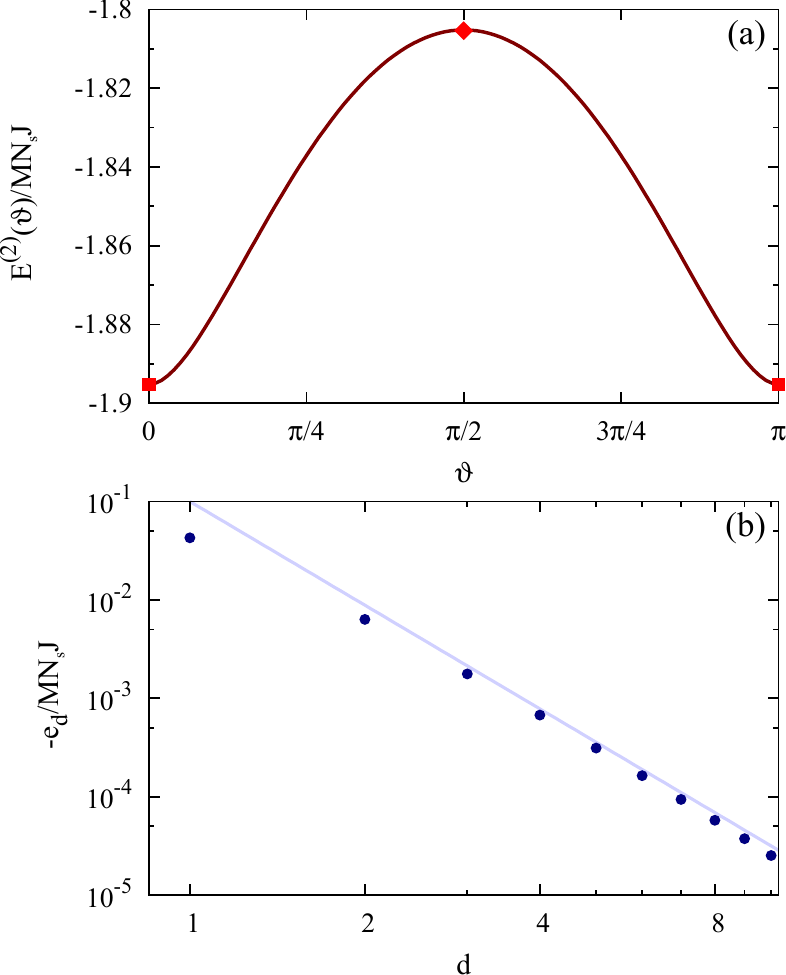}
\caption{(a) Quantum correction to the energy of the helical state as a function of $\vartheta$. The energy takes its minimum value for $\vartheta=0$ and $\vartheta=\pi$, which correspond to the four-sublattice state shown in Fig.~\ref{fig:trivial}. The red squares and the diamond correspond to the values given in Eqs.~(\ref{eq:energy_theta_nul}) and (\ref{eq:energy_theta_pio2}), respectively.
 (b) The coefficients $ e_d$ of the Fourier series of the energy, defined as $E^{(2)}(\vartheta)/N_s = e_0 + \sum_{d=1}^{\infty} e_d \cos 2 d \vartheta$, plotted on a log-log scale. The first two coefficients are $e_0 = -1.843 M J $ and $e_1 = -0.043 M J$. Further coefficients scale with the index $d$ as $\propto d^{-3.5}$ (straight line) showing a nonanalytical behavior at $\vartheta=0$. 
 }
\label{fig:energy} 
\end{figure}

After diagonalization, the correction to the ground state energy per site evaluates to

\begin{equation}
\frac{E^{(2)}(\vartheta)}{N_s} =
\frac{1}{2}  \sum_{i=1}^6 \left[\int_{\text{BZ}} \frac{\text{d}^3 \mathbf{q}}{32\pi^3} \, \frac{\omega_i(\mathbf{q})}{2} - 4 JM \right],
\label{Eq:energy}
\end{equation}
where $32\pi^3$ is the volume of the Brillouin zone shown in Fig.~\ref{fig:ordering_Q}(a) and the factor 1/2 in front of the integral takes into account that our unit cell has two sites. The correction is an even function of $\vartheta$, $E^{(2)}(\vartheta)=E^{(2)}(-\vartheta)$  and is periodic with period $\pi$, $E^{(2)}(\vartheta)=E^{(2)}(\pi+\vartheta)$ [while the $\vartheta \to \vartheta+2\pi$ leaves the configurations in the helical state unaltered, the periodicity in $\pi$ is less obvious: $\vartheta \to \vartheta+\pi$ exchanges the $C$ and $D$ basis states, see Eqs. \eqref{eq:xi_recursive} and \eqref{eq:R}, which can be undone by a global SU(4) rotation].
The ground state energy (\ref{Eq:energy}) is plotted for $0\leq \vartheta \leq \pi$ in Fig.~\ref{fig:energy}. The lowest energy configuration is realized for $\vartheta = 0$. Thus, this analysis shows, that among the helical states, the ground state is the one with $\vartheta=0$, which is a non-helical, simple, four-sublattice state. 

\subsection{The $\vartheta = 0$ ground state}
\label{ssec:thet0gs}
For a general $\vartheta$, the characteristic equation of the Bogoliubov matrix, Eq.~\eqref{eq:bogodiag}, can be evaluated only numerically. However, in the $\vartheta = 0$ and $\vartheta = \pi/2$ cases
it becomes analytically solvable. Here we illustrate this property for $\vartheta = 0$ (the $\vartheta=\pi/2$ case is treated in Appendix~\ref{app:pi_2_case}).

In this case the Bogoliubov matrix, Eq.~(\ref{eq:M_mtx}), can be transformed into a block diagonal matrix, where the blocks are all of dimension $2\times2$, and are all of the same general form
\begin{equation}
\mathbf{M}_{\nu}^{\text{sq}} =
4 JM \, \left[
 	\begin{array}{cc}
	1 & \gamma_{\nu}^{\textrm{sq}}(\mathbf{q}) \\
	-\gamma_{\nu}^{\textrm{sq}}(\mathbf{q}) & -1
 	\end{array}
\right] \, , \quad\nu \in \{ x,y,z \},
\label{eq:msq}
\end{equation}
where
\begin{subequations}
\label{eqs:gammas}
\begin{eqnarray}
\gamma^{\textrm{sq}}_{x}(\mathbf{q}) &=& \cos \frac{q_y}{2} \cos\frac{q_z}{2} \, , \\
\gamma^{\textrm{sq}}_{y}(\mathbf{q})  &=&\cos\frac{q_x}{2} \cos\frac{q_z}{2} \, , \\
\gamma^{\textrm{sq}}_{z}(\mathbf{q})  &=&\cos\frac{q_x}{2} \cos\frac{q_y}{2} \, . 
\end{eqnarray}
\end{subequations}
The only difference between the blocks is in $\gamma_{\nu}^{\textrm{sq}}(\mathbf{q})$. All three different blocks, with different $\nu$, participate twice in the matrix (\ref{eq:M_mtx}).

This simplification is the consequence of the highly symmetric, four-sublattice structure of this phase. To be more specific, there is a single flavor on each site, i.e. the  mean value of the Schwinger boson operators is one of the basis vectors: $ \bm{\xi}_A$, $\bm{\xi}_B$, $\bm{\xi}_C$, or $\bm{\xi}_D $.
There is a one-to-one correspondence between a pair of colors: $AB$, $AC$, $AD$, $BC$, $BD$, $CD$, and a specific plane characterized by its normal vector [one of $(1,0,0)$, $(0,1,0)$, or $(0,0,1)$] and its parity (whether it has integer or half-integer coordinates along the direction of the normal vector). Inside each plane, the corresponding specific two colors alternate in a checkerboard manner. The interaction term in the Hamiltonian between a pair of neighboring sites involves the product of two Holstein-Primakoff bosons. One of them is from the first and the other one is from the other site. The Holstein-Primakoff boson on the first site has to be from the same flavor as the classical ordering of the other site, and vice versa in order to survive the contraction of the indices in Eq.~\eqref{eq:permutation} -- see also Appendix A1 in Ref.~[\onlinecite{Corboz_honeycomb_2012}]. Therefore, for a given flavor pair, there is a specific plane out of the $6=3\times2$ different ones (3 for the directions and 2 for the parity) within which the given type of boson flavors can interact. 
Such an interaction is described by one of the $\mathbf{M}_{\nu}^{\text{sq}}$ matrices.
In fact, for the state depicted in Fig.~\ref{fig:trivial}, the $A$ and $B$ boson interact only within the planes characterized by integer $z$ values, the $C$ and $D$ bosons interact within planes with half-odd integer $z$ values, the $A$ and $D$ bosons within the plane with integer $x$ values and so on.

\begin{figure}[t]
\includegraphics[width=0.9 \columnwidth]{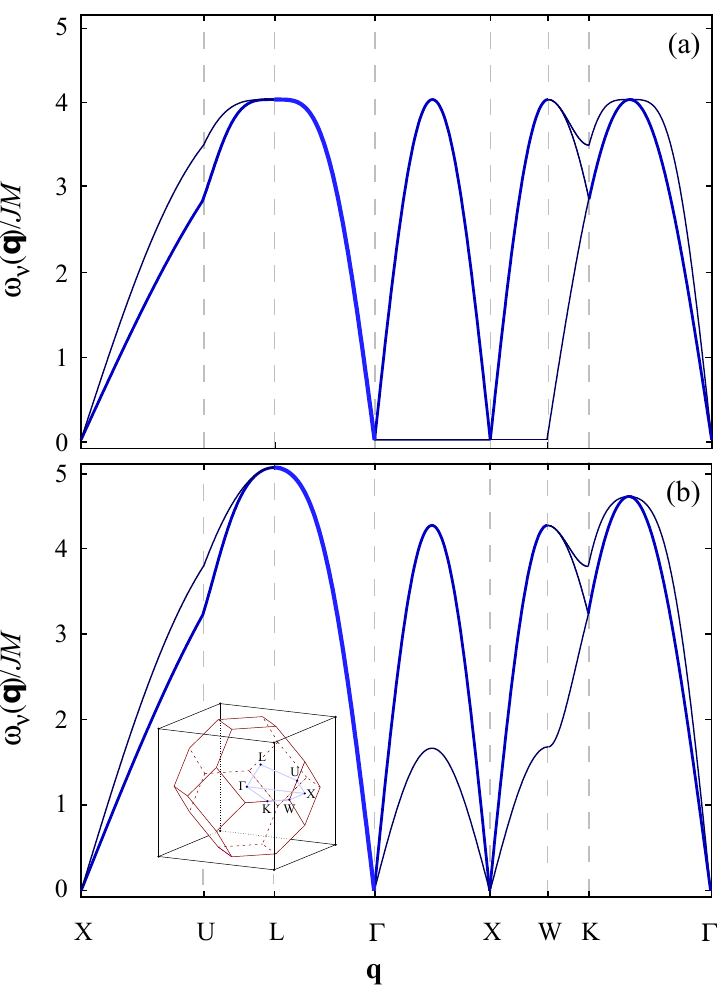}
\caption{(color online) Dispersion relation of the flavor waves for the Heisenberg model with nearest neighbour interactions only (a), and with a finite second neighbour ferromagnetic exchange $J_2/J=-0.086$ (b), along the path in the Brillouin zone shown in the inset. The degeneracy ($\times1$, $\times2$, and $\times3$) of each mode is proportional to the width of the lines and the colour becomes lighter with increasing degeneracy.
 \label{fig:spagetti}}
\end{figure}

Since these parts of the Hamiltonian are  independent, they can be diagonalized separately. 
The excitation energies of the above $\mathbf{M}_{\nu}^{\text{sq}}$ matrices are 
\begin{equation}
\omega_\nu^{\textrm{sq}}(\mathbf{q}) = 4 JM\sqrt{1-\left| \gamma_\nu^{\text{sq}}(\mathbf{q}) \right|^2} \,.
\label{eq:spect_theta_0}
\end{equation}
 The dispersion $\omega_{\nu}(\mathbf{q})$ is flat along the $\nu$ direction, as $q_\nu$ is missing from the corresponding $\gamma^{\textrm{sq}}_{\nu}(\mathbf{q})$ in Eqs.~\eqref{eqs:gammas}. The dispersions are shown in Fig.~\ref{fig:spagetti}(a).

The contribution to the single site energy from the different $\mathbf{M}_{\nu}^{\text{sq}}$ blocks give the same value:
\begin{equation}
\frac{E^{(2)}_{\text{sq}}}{N_s} = - 0.316 JM \, .
\end{equation}
Adding all these contributions, the total quantum correction to the energy, Eq.~(\ref{Eq:energy}), evaluates to
\begin{equation}
\frac{E^{(2)}_{\vartheta = 0}}{N_s} = 6  \frac{E^{(2)}_{\text{sq}}}{N_s}  = -1.896 \, JM\, .
\label{eq:energy_theta_nul}
\end{equation}

We draw the attention to the flat, zero energy mode between the $\Gamma$ and X points in the Brillouin zone. It is the Goldstone mode associated with the continuous classical degeneracy of the ground state,\cite{goldstone1962broken} namely that each plane can be independently rotated in the SU(4) space, while keeping the two sublattice structure within the plane, as expressed by the freedom of choosing the $\vartheta$'s in Eq.~(\ref{eq:vartheta_Z}).

\section{Hamiltonian extended with second neighbour exchange}
\label{sec:NNN}

In the case of the SU(2) spin systems, quantum fluctuation prefer collinear structures, thus it is customary to approximate the effect of quantum and thermal excitations by adding a {\it biquadratic} term with a negative coefficient of the order of $1/S$ to the classical energy, $S$ being the length of the SU(2) spin.\cite{Henley1987,henley1989ordering} The classical solution of such an effective model reproduces the phenomena attributed to quantum fluctuations, among others, the magnetisation plateaus on frustrated lattices (see Ref.~\onlinecite{Zhitomirsky2015} for a more detailed discussion). In the case of SU(4) spins, however, the notion of the collinearity is not so obvious (for example, adding na\"ively a $\left| \sum_{\alpha} \xi_{\mathbf{r},\alpha}^* \xi^{\phantom{*}}_{\mathbf{r}',\alpha} \right|^4$ term to the expression of classical energy, Eq.~(\ref{eq:H_exp}), does not help in the selection of the ground state). Furthermore, the nature and origin of the degeneracy we need to reduce is different in the SU(2) and SU(4) case: while in the frustrated SU(2) models the ground states do not optimise the bond energies, in the SU(4) case each bond is fully satisfied at the classical level. 

A hint is given by the observation that the second (i.e. next nearest) neighbour ferromagnetic interaction is compatible with $\vartheta=0$ four sublattice order selected by the fluctuations. For this reason, we extend the Hamiltonian with the exchanges over the $\left\langle \left\langle \mathbf{r},\mathbf{r'}  \right\rangle \right\rangle$ next-nearest-neighbor pairs
\begin{eqnarray}
\mathcal{H}^{\text{ext}} &=& \mathcal{H}  +  J_2 \sum_{\left\langle \left\langle \mathbf{r},\mathbf{r'}  \right\rangle \right\rangle}  \sum_{\alpha, \beta} S^\alpha_\beta(\mathbf{r}) S_\alpha^\beta(\mathbf{r}') \, .
\label{eq:H_eff}
\end{eqnarray}
Each site has 12 nearest neighbour and 6 second neighbour sites. 
We note in passing that antiferromagnetic $J_2>0$ interactions lead to frustration in the usual sense: we are not able to satisfy all the bonds at the same time. Below, we restrict the discussion of the spectrum to the antiferromagnetic $J$ and ferromagnetic $J_2$, in order to be compatible with the $\vartheta = 0$ ordering, which is realized for any $J>0$ and $J_2<0$ values. The classical energy of the  ground state is then 
\begin{align}
  \frac{E^{(0)}}{N_s} = 3 J_2 M^2, 
\end{align}
while the expectation value of the Hamiltonian (\ref{eq:H_eff}) in the helical state is  
\begin{align}
  \frac{E^{(0)}(\vartheta)}{N_s} = \left(3 -\sin^2 \vartheta\right) J_2 M^2 \;. 
\end{align}
Comparing this with the energy of the quantum fluctuations, in particular with Fig.~\ref{fig:energy}(b), we get that the fluctuations can be described by choosing 
\begin{equation}
J_2^{\mathrm{eff}} = \frac{2 e_1}{M^2} = -0.086\frac{J}{M} 
\label{eq:J2eff}
\end{equation}
 at the classical level. In analogy to the SU(2) case,\cite{yildirim1998,yildirim1999} we anticipate that the addition of this effective coupling will provide a finite, $\propto 1/M^2$, dispersion to the zero mode between $\Gamma$ and $X$.

To get the dispersion with finite $J_2$, we repeat below the calculation presented in Sec. \ref{sec:quantum}, now with the inclusion of the $J_2$ coupling. The $J_2$ does not bring in any further complication, as it simply couples the Holstein-Primakoff bosons in different layers, so that the Bogoliubov matrix \eqref{eq:msq} is replaced by
\begin{equation}
\mathbf{M}_{\nu}^{\text{ext}} = \mathbf{M}_\nu^{\text{sq}} + \mathbf{M}^{\text{NNN}}  \, , 
\label{eq:ext_bogo}
\end{equation}
where $\mathbf{M}_\nu^{\text{sq}}$ is still given by Eq.~(\ref{eq:msq}), and 
\begin{equation}
\mathbf{M}^{\text{NNN}}  = - 6 J_2 M  \left[
 	\begin{array}{cc}
	\bar\gamma^{\text{NNN}}(\mathbf{q}) &  0  \\
	0 & -\bar\gamma^{\text{NNN}}(\mathbf{q})
 	\end{array}
\right] \, ,
\label{eq:bogo_ferro}
\end{equation}
with
\begin{equation}
\label{eq:dispnnn}
\bar\gamma^{\text{NNN}}(\mathbf{q}) = 1-\frac{\cos q_x+\cos q_y+\cos q_z}{3} \, .
\end{equation}
This modified Bogoliubov matrix has the following spectrum:
\begin{equation}
\omega_\nu^{\textrm{ext}}(\mathbf{q}) = M \sqrt{[4 J - 6 J_2 \bar\gamma^{\text{NNN}}(\mathbf{q})]^2- \left[ 4 J  \gamma_\nu^{\text{sq}}(\mathbf{q}) \right]^2} \, .
\label{eq:eff_spect}
\end{equation}

The spectrum, depicted in Fig.~\ref{fig:spagetti}(b) for above mentioned effective value $J_2/J=-0.086$, is modified from that of the $J_2=0$ case of Fig.~\ref{fig:spagetti}(a), most noticable along the path connecting the $\Gamma$ and $X$ points, where the spectrum obtains a dispersion now. It is interesting to observe, and also useful for testing the consistency of our calculation, that the ferromagnetic nature of the next-nearest-neighbor coupling is manifested in the diagonal part of the Bogoliubov matrix, Eq.~(\ref{eq:bogo_ferro}). 
Thus, in the case of $J\ll |J_2|$ the dispersion $\omega_\nu \propto |\mathbf{q}|^2$, as it is expected for a ferromagnetic system, except in the close vicinity of  the $\Gamma$ point, where the dispersion remains linear with $|\mathbf{q}|$ for $|\mathbf{q}| \lesssim \sqrt{J/J_2}$

\section{Spin reduction and ordering at zero and finite temperatures}
\label{sec:correction}

The reduction of the magnetization, $\mu_{\mathbf{r},\alpha}$ defined by Eq.~\eqref{eq:spinred}, is a measure of how good our approximation is. In the semiclassical expansion, Eq.~(\ref{eq:aprox}) in particular, we assume that $\mu_{\mathbf{r},\alpha}\ll M$. Therefore, for $M=1$ one needs the reduction to be much smaller than unity. More permissively, e.g. for allowing higher values of $M$, one can require $\mu_{\mathbf{r},\alpha}$ to be at least finite. In this section, we calculate the reduction for zero and finite temperatures for the four sublattice state.

\begin{figure}[bt]
\includegraphics[width=0.9 \columnwidth]{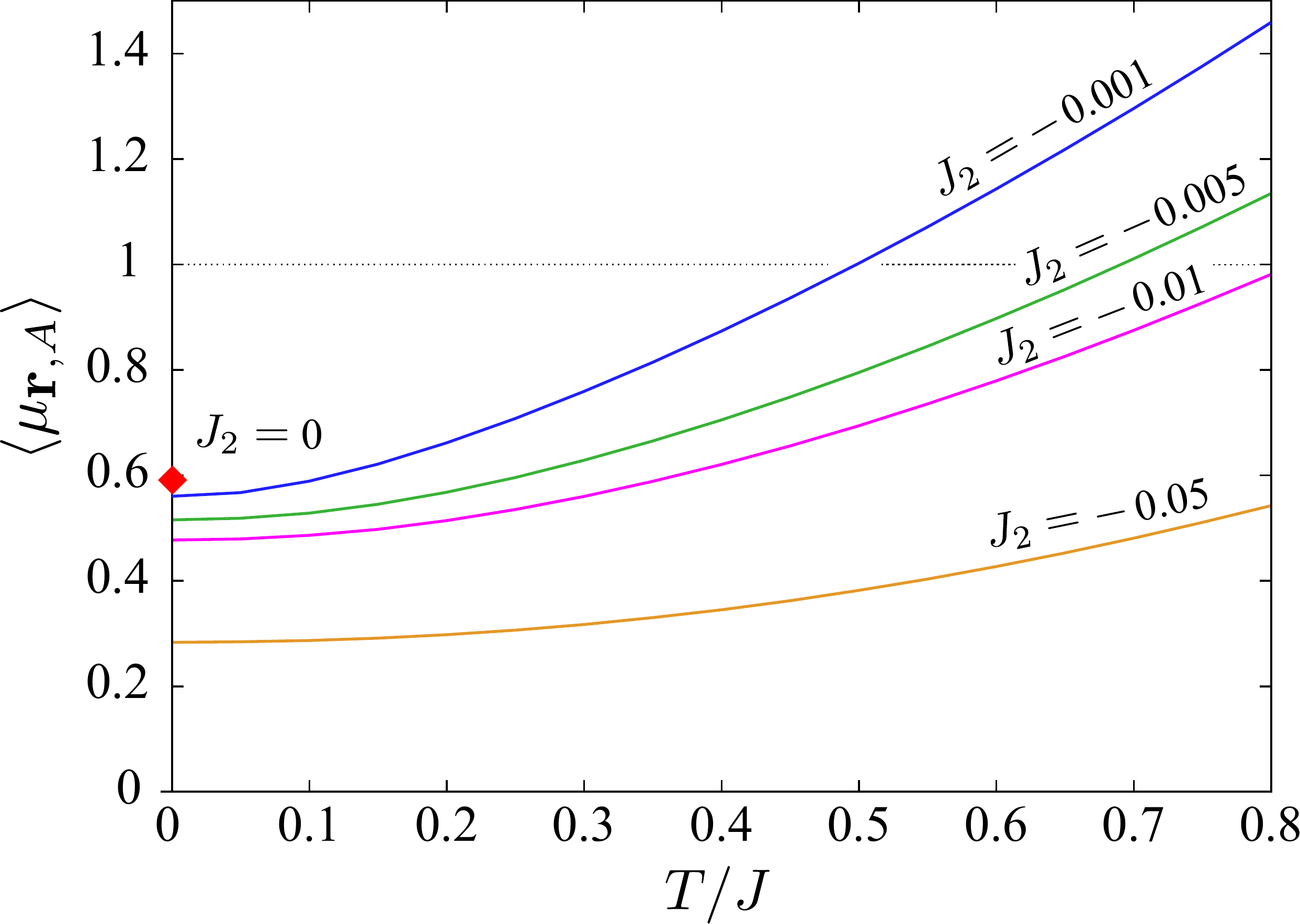}
\caption{(color online) The spin reduction vs temperature for different values of the ferromagnetic, next-to-nearest-neighbor coupling $J_2$. For $J_2=0$ the spin reduction is finite only at zero temperature, otherwise it diverges. For nonzero $J_2$ the spin reduction is finite also at nonzero temperatures. The horizontal gray line at $\mu_{\mathbf{r},A}=1$ indicates an upper limit of the model's applicability.}
 \label{fig:spinred}
\end{figure}

The mean value of the spin on sublattice $\Lambda_A$ reads as,
\begin{equation}
\left< S_A^A(\mathbf{r}) \right> = M  - \left< \mu_{\mathbf{r},A} \right> ,\quad \mathbf{r}\in\Lambda_A,
\label{eq:spin_red}
\end{equation}
and an equivalent expression holds for $\left< S_B^B(\mathbf{r}) \right>$ on the $\Lambda_B$ sublattice. The reduction can be expressed with the Fourier-transformed boson vector, 
\begin{equation}
\left<\mu_{\mathbf{r},A}\right>=\!\!\!\sum_{\alpha \in\lbrace B,C,D\rbrace}\sum_{m = 1}^{12}  \int_{\text{BZ}} \frac{\text{d}\mathbf{q}}{32 \pi^3} T_{\alpha,m}^{*}(\mathbf{q}) T_{\alpha,m}^{\phantom{*}} (\mathbf{q}) \, C_m(\mathbf{q}) \, ,\label{eq:spin_reduction}
\end{equation}
where
\begin{equation}
C_m(\mathbf{q})=
\begin{cases}
\left< \tilde{\mathbf{\Phi}}_{m}^\dagger (\mathbf{q}) \tilde{\mathbf{\Phi}}_{m}^{\phantom{\dagger}} (\mathbf{q}) \right>,&\text{for }m=1,\ldots,6,\\
1+\left< \tilde{\mathbf{\Phi}}_{m}^\dagger (\mathbf{q}) \tilde{\mathbf{\Phi}}_{m}^{\phantom{\dagger}} (\mathbf{q}) \right>,&\text{for } m=7,\ldots,12.
\end{cases}
\end{equation}

\subsection{Spin reduction at zero temperature}

At zero temperature the quasiparticle occupation is zero, therefore we have $C_m(\mathbf{q})=1$ only for $m=7,\ldots,12$.
For the four-sublattice state and for $J_2=0$ Eq.~(\ref{eq:spin_reduction}) can be evaluated to 
\begin{align}
\label{eq:spin_red_0}
\left<\mu_{\mathbf{r},A}\right> &= \frac{1}{2} \sum_{\nu} \int_{\text{BZ}} \frac{\text{d}^3\mathbf{q}}{32 \pi^3} \left[ \frac{1}{\sqrt{1-\left| \gamma_\nu^{\text{sq}}(\mathbf{q}) \right|^2}} - 1 \right] \nonumber \\
&\approx 0.590 \, .
\end{align}
It is noteworthy to compare the result of the spin reduction with that of a single square lattice. In our case the spin reduction is three times the value of that of the square lattice case, which was found to be approximately 0.197.\cite{fazekas1999lecture} The factor 3 can be understood as the contribution of the 3 Holstein-Primakoff bosons describing fluctuations in one of the three orthogonal square lattice crystallographic planes. The small finite spin reduction gives a consistent result with our initial assumption of the ground state being ordered. At the same time, this value is quite large for a three dimensional magnet, and the reason is the flat dispersion between the $\Gamma$ and X points in the Brillouin zone [Fig.~\ref{fig:spagetti}(a)] and thus the reduction of the dimensionality from 3 to 2.

For a finite $J_2<0$, fluctuations are interacting along all 3 directions, and the number of average bosons decreases rapidly with the increase of the ferromagnetic $J_2$, as seen in Fig.~\ref{fig:spinred} at the $T=0$ point. For example, a small $J_2/J = -0.01$ reduces the fluctuations to $0.42$, and for the $J_2^{\mathrm{eff}} = 0.086J$ [taking $M=1$ in Eq.~(\ref{eq:J2eff})] the $\left< \mu_{\mathbf{r},A} \right> \lesssim 0.1$.

\subsection{Spin reduction at finite temperatures}

At low but finite temperature the Holstein-Primakoff bosons obtain thermal occupation. The spin reduction now contains contributions both from quantum and thermal fluctuations 
\begin{align}
 \left<\mu_{\mathbf{r},A}\right> = \sum_{\nu=x,y,z} \int_{\text{BZ}} \frac{\text{d}^3\mathbf{q}}{32 \pi^3} \left\{ \frac{4JM }{\omega_\nu^{\textrm{ext}}(\mathbf{q})} \left[ n_\nu({\mathbf{q})} +\frac{1}{2}\right] - \frac{1}{2} \right\}  \label{eq:spin_red_T}
\end{align}
where 
\begin{align}
n_\nu(\mathbf{q}) = \frac{1}{ \exp(\omega^{\mathrm{ext}}_\nu(\mathbf{q}) /T)-1} 
\end{align}
is the occupation number of the bosonic excitations.

For $J_2=0$ the integral evaluates to finite value only at the $T=0$ point, its value given in Eq.~\eqref{eq:spin_red_0}. At finite temperature, thermal occupation of the Goldstone mode diverges along the line where the spectrum, Eq. \eqref{eq:spect_theta_0}, is constant zero. As a consequence, the integral becomes divergent for $T>0$. This is in accordance with the Mermin-Wagner-Hohenberg theorem \cite{mermin1966absence,hohenberg1967existence}, as the effective dimensionality of the problem in the linear flavor wave approximation is 2 when $J_2=0$. 

For finite $J_2$, the  integral~\eqref{eq:spin_red_T} with the dispersion \eqref{eq:dispnnn}, yields a finite spin reduction even at finite temperatures. We plot the value $\langle\mu_{\mathbf{r},A}\rangle$ for different values of $J_2$ in Fig.~\ref{fig:spinred}. For a smaller value of $|J_2|$, the value of the integral is larger, and is more sensitive to the temperature.  Thus the small but always present longer range interactions stabilize the four-sublattice-ordered state at finite temperature. In the case of the quantum model, the effective $J_2^{\mathrm{eff}}$ generated by quantum fluctuation plausibly stabilizes the long range ordered state, with a transition temperature that might substantially deviate from a mean-field like estimate.

\section{Discussion}
\label{sec:sum}

We  investigated spin ordering in the case of an SU(4) symmetric Heisenberg model on an fcc lattice. 
For this purpose, we used the Schwinger boson representation of the SU(4) spin operators, which allows a controlled approximation of the quantum effects in an SU(4) symmetry-broken state with long range order.

At the classical level, where all Schwinger bosons are replaced by their expectation values, the ordered state was found to be highly degenerate, consisting of layers of two-sublattice ordered square lattice planes, however the states on the neighboring planes are determined up to SU(2) rotations.

Considering a subset of these degenerate states using the flavor-wave theory, we found that quantum fluctuations select a four sublattice ordered state, with a magnetic unit cell being identical to the cubic unit cell of the fcc lattice. 
The linear flavor wave Hamiltonian of the four sublattice ordered state block diagonalizes into a set of two-flavor bosons interacting on decoupled planes, so that fluctuations couple strongly only along two directions. This effect renders the ordering fragile at finite temperatures, as the number of bosons diverge at any $T>0$. 
Coupling between the planes is established by a ferromagnetic second neighbor spin interaction, which stabilizes the four-sublattice classical ordering up to some finite ordering temperature. 
This second neighbor spin interaction can be provided by the quantum fluctuation themselves, since they also favor the four sublattice state. It would be interesting to see how does the N\'eel temperature in this case relate to the mean-field result (i.e. to the Curie-Weiss temperature).

It is noteworthy to compare these result with the usual, SU(2) symmetric Heisenberg model on the same fcc lattice. In both cases the classical ground state is highly degenerate, however the nature and origin of the degeneracy are different.
Since the fcc lattice is composed of edge-sharing tetrahedra, the SU(2) classical spins -- represented by three dimensional vectors -- are frustrated and cannot satisfy all the bonds optimally (unlike to the SU(4) case). The lowest energy state on a single tetrahedron is achieved when the sum of the four spins on the four corners adds up to 0, which can be realized in continuously many ways, leading to a high degeneracy also on the fcc lattice. The idea of the selection of the ordered state from the degenerate classical manifold by minimizing the energy of the quantum fluctuation was pioneered for this lattice: 
Quantum fluctuation were shown to reduce the continuous degeneracy to discrete ones, as they generally favor collinear configuration, in which case the O(3) spins can be treated as Ising spins.\cite{shender1982,henley1989ordering} 
The effect of quantum fluctuations for SU(2) Heisenberg models including longer range interactions or anisotropies has been extensively studied in case of type I\cite{Oguchi1985,Henley1987,yildirim1998,Ader2001} or type II\cite{Datta2012} orderings. However, even after many efforts, the true nature of the ground state of the for the SU(2) Heisenberg model with nearest neighbor exchanges is still unknown.  

We note that there is a fundamental difference between the SU(4) and SU(2) spins on the fcc lattice: for the SU(2) case the number of constraints is larger than the spin degrees of freedom, thus the bond energies cannot be all satisfied. In contrast, for the SU(4) spins there are more degrees of freedom as constraints, and the ground state is degenerate with all the bond energies fully satisfied.

\begin{acknowledgments}
This work was supported by the Hungarian OTKA Grants No. K106047 and PD104652,
and the Hungarian Academy of Sciences (Lend\"ulet Program, LP2011-016). G.~Sz. also acknowledges support from the Bolyai J\'anos Fellowship.
\end{acknowledgments}

\appendix 
\section{Zero point energy for $\vartheta=\pi/2$}
\label{app:pi_2_case}
\begin{figure}[t]
\includegraphics[width=0.7 \columnwidth]{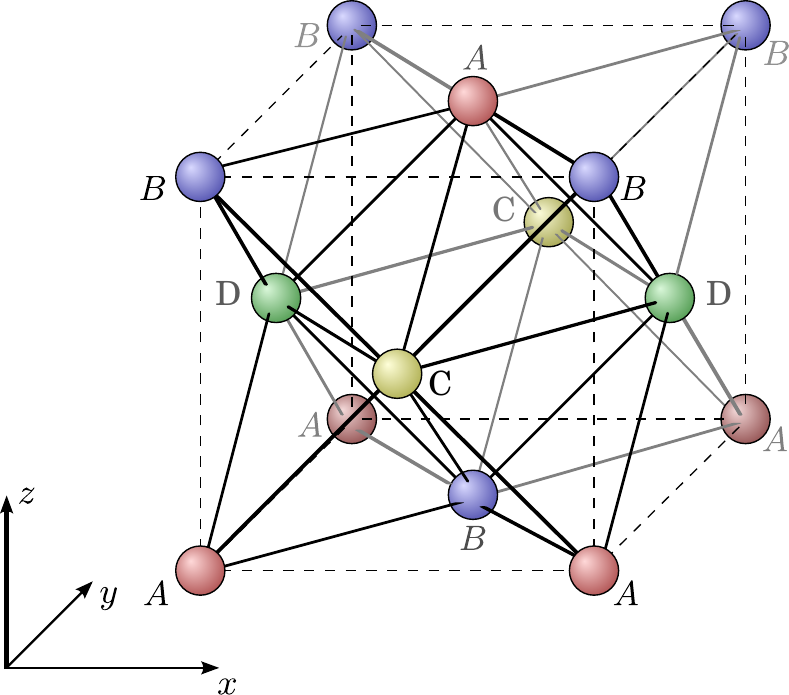}
\caption{(color online) The twisted four-sublattice state. The coloring of the sites on the figure corresponds to $\vartheta = \pi/2$ in Eq.~(\ref{eq:rotatingfield}).
 \label{fig:theta_p_2}}
\end{figure}

Here, we calculate the quantum correction to the energy for $\vartheta = \pi/2$. This ansatz is shown in Fig.~\ref{fig:theta_p_2}.  Two of the Holstein-Primakoff boson pairs now make square lattices (the red--blue and green--yellow planes), while the remaining pairs of colors make a lattice isomorphic to the diamond lattice (for example, the red--yellow). The Bogoliubov matrix Eq.~(\ref{eq:M_mtx}) can be transformed to a block diagonal form by making a symmetric and antisymmetric combination of certain bosons in $\mathbf{\Phi}$.  Each block is a $2\times2$ matrix and two of them have the same form as in the $\vartheta = 0$ case with $\gamma_z^{\text{sq}}$, and the other four have the following form:
\begin{equation}
	\mathbf{M}^{\textrm{diam}}(\mathbf{q}) = 
 	4  JM \left[
 	\begin{array}{cc}
	1 & \gamma^{\textrm{diam}}(\mathbf{q}) \\
	-\gamma^{\textrm{diam}}(\mathbf{q}) & -1
 	\end{array}
 	\right] \, ,
\end{equation}
where
\begin{eqnarray}
\gamma^{\textrm{diam}}(\mathbf{q}) &=& \frac{1}{4}\left( e^{i \frac{1}{4}(q_x+q_y+q_z)} + e^{i \frac{1}{4}(q_x-q_y-q_z)} \right. \nonumber \\
&& + \left. e^{i \frac{1}{4}(q_y-q_x-q_z)} + e^{i \frac{1}{4}(q_z-q_x-q_y)}\right) \, .
\end{eqnarray}
This Bogoliubov matrix yields
\begin{equation}
E^{(2)}_{\text{diam}} = - 0.293 JM \, ,
\end{equation}
which modifies the total energy correction to
\begin{equation}
E^{(2)}_{\vartheta = \pi/4} =  2 E^{(2)}_{\text{sq}} + 4 E^{(2)}_{\text{diam}}  = -1.804 \, JM  \, .
\label{eq:energy_theta_pio2}
\end{equation}
This energy is higher than that of the $\vartheta = 0$ case Eq.~(\ref{eq:energy_theta_nul}). Actually, this configuration has the highest energy of all the helical states. Its value is illustrated with the green diamond in Fig.~\ref{fig:energy}(a).

\bibliography{magnetism}

\end{document}